\documentclass [12pt]{article}
\usepackage {graphicx}

\providecommand{\keywords}[1]
{
	\small	
	\textbf{\textit{Keywords---}} #1
}
\title{Parameter space arrangement \\ for a model system nearby domain of existence \\ of
Plykin type attractor}

\author{Kuznetsov S.P. and Sataev I.R. \\
\small Kotel'nikov Institute of Radioengineering and Electronics of RAS, Saratov Branch, \\
\small Zelenaya 38, Saratov, 410019, Russian Federation
}

\begin{document}
\maketitle

\begin{abstract}
For a model system defined as combination of sequentially applied continuous
transformations of a sphere, the question of arrangement of the parameter
space around the domain of existence of the Plykin-type attractor is
considered. Results of numerical calculations are presented, including
charts of dynamical regimes and Lyapunov exponents on the parameter plane,
as well as portraits of attractors in characteristic regions of chaotic and
regular dynamics. The Plykin attractor region is determined and depicted
using the computational procedure for checking hyperbolicity, which consists
in analyzing angles between expanding and contracting tangent subspaces of
typical trajectories on the attractor. The Plykin attractor takes place in a
bounded continuous region of the parameter plane that corresponds to
roughness (structural stability) of the hyperbolic dynamics. Outside that
region, various types of dynamics are observed including non-hyperbolic
chaos, periodic and quasiperiodic behaviors.
\end{abstract}







\newpage

Uniformly hyperbolic attractors introduced in the second half of the 20th
century in a framework of rigorous mathematical theory provide examples of
strictly justified deterministic chaos in dynamical systems [1,2,3]. These
are attractors consisting exclusively of saddle-type phase trajectories, so
that each trajectory has stable and unstable manifolds formed by sets of
neighboring orbits approaching or moving away from the reference trajectory.

Until recently, only artificial mathematical constructions like
Smale--Williams solenoid and Plykin attractor were considered as examples of
the uniformly hyperbolic attractors. The Smale--Williams attractor appears
while mapping a toroidal domain in the phase space of minimal dimension
3 with longitudional expansion and transversal contracting into itself
been being rolled up into a multi-turn loop. The Plykin attractor [4,5] can
be realized for a map on a sphere with four holes, or in a bounded region on
a plane with three holes.

A fundamental mathematical fact is that the hyperbolic chaos has the
property of roughness, or structural stability, which consists in the fact
that it exhibits dynamics, which do not change qualitatively under a small
variation of parameters [6]. This property would be of exceptional
importance for natural systems and for technical applications, providing
insensitivity of chaos characteristics to inaccurate parameter settings,
manufacturing errors, various noises and disturbances.

Recently, examples of systems admitting a physical realization have been
proposed, where the Smale--Williams or Plykin hyperbolic attractors take
place for the mappings, which arise when constructing the Poincar\'e sections
[7].

A fundamentally important question in the search, design, and use of systems
with hyperbolic chaos is how the dynamics depend on parameters in regions where
occurrence of such chaos is potentially possible. Theoretically,
this is valuable for enriching mathematical concepts, as a step from
considering rough situations to situations of varying degrees of non-roughness,
like in the traditional theory of bifurcations. Practically, this is
important for optimizing characteristics of the designed chaos generators.
Also, information about the arrangement of the parameter space surrounding
the area of hyperbolic chaos can help in the search for systems of natural
origin, which manifest the hyperbolic dynamics (mechanics, neurodynamics, chemical
kinetics, etc.).

By virtue of roughness, hyperbolic chaos must occupy continuous domains in
the parameter space, but the environment structure of these regions is of
interest as characterizing bifurcation scenarios of transition to the
hyperbolic chaos. The available literature on this issue so far refers only
to the simplest type of hyperbolic attractors, namely, the Smale--Williams
solenoids [9, 10, 11, 12].

This article discusses arrangement of the parameter space region surrounding
the area of existence of the Plykin attractor in the model system [8]. The
system is constructed using a certain periodic sequence of continuous
transformations of the sphere. To make the task more meaningful, we first
modify the model, supplementing it with parameters, depending on which the
analysis of dynamical behavior will be conducted. Further, we present
results of computations, including visualization of charts of dynamical
regimes and Lyapunov exponents on the parameter plane, as well as attractor
portraits in characteristic regions of chaotic and regular dynamics. In a
frame of these studies, the region of existence of the Plykin hyperbolic
attractor is determined using the special computational procedure for
checking hyperbolicity, based on the analysis of angles between the
expanding and contracting tangent subspaces of typical trajectories on the
attractor [15-17].

\section{The model system}
\label{subsec:mylabel1}

\bigskip

Following [8], we assume that for our system instantaneous states are given
by points of the unit sphere, which can be represented in angular
coordinates $(\theta ,\,\varphi )$ or in rectangular coordinates
$(x,\,y,z)$:

\begin{equation}
\label{eq1}
x = \cos \varphi \sin \theta ,
\quad
y = \,\sin \varphi \sin \theta ,
\quad
z = \cos \theta ,
\end{equation}
where $x^2 + y^2 + z^2 = 1$ (Fig.1). As proved by Plykin, a map of a sphere
onto itself can have a hyperbolic attractor only in presence of at least
four holes, the regions not visited by trajectories on the attractor. In our
construction the holes correspond to neighborhoods of the points A, B, C, D
with coordinates $(x,y,z) = (\pm 1 / \sqrt 2 ,\,\,0,\,\,\pm 1 / \sqrt 2 )$,
see Fig.1. The North and South poles of the sphere are denoted as N and S,
respectively.

Let us consider a sequence of four continuous transformations:

I. Displacement of representative points per unit time along the parallels
from the meridians NABS and NCDS towards the meridian circle equidistant from
them:

\begin{equation}
\label{eq2}
\dot {x} = - \varepsilon xy^2,\,\,\,\dot {y} = \varepsilon x^2y,\,\,\,\,\dot
{z} = 0.
\end{equation}

II. Differential rotation per time $T_{1 }$ around $z$ axis with angular velocity
linearly dependent on $z$:

\begin{equation}
\label{eq3}
\dot {x} = \pi (z / \sqrt 2 + 1 / 2)y,\,\,\,\,\dot {y} = - \pi (z / \sqrt 2
+ 1 / 2)x,\,\,\,\,\dot {z} = 0.
\end{equation}

III. Displacement of representative points per unit time on the sphere along
circles with centers at the axis $x$ from a large circle ABCD towards the equator:

\begin{equation}
\label{eq4}
\dot {x} = 0,\,\,\,\dot {y} = \varepsilon yz^2,\,\,\,\dot {z} = -
\varepsilon y^2z.
\end{equation}

IV. Differential rotation per time $T_{2 }$ around $x$ axis with angular velocity
linearly dependent on $x$:

\begin{equation}
\label{eq5}
\dot {x} = 0,\,\,\,\dot {y} = - \pi (x / \sqrt 2 + 1 / 2)z,\,\,\,\dot {z} =
\pi (x / \sqrt 2 + 1 / 2)y.
\end{equation}

Unlike the previously considered model, the durations of the stages II and
IV are considered as parameters, depending on which the dynamical behavior
analysis will be carried out, whereas in [8] they were fixed
($T_{1}=T_{2}$=1).

The Poincar\'{e} map ${\rm {\bf x}} \mapsto {\rm {\bf F(x)}}$ describing
evolution of the system over a period $T = 2 + T_1 + T_2 $ is the result of
sequential application of four mappings, corresponding to the stages
I-IV, to the initial state ${\rm {\bf x}} = (x,y,z)$:

\[
{\rm {\bf x}} \mapsto \left( {\begin{array}{l}
 x\sqrt {\frac{x^2 + y^2}{x^2 + y^2e^{2\varepsilon (x^2 + y^2)}}} ,\, \\
 y\sqrt {\frac{x^2 + y^2}{x^2e^{ - 2\varepsilon (x^2 + y^2)} + y^2}} ,\, \\
 \,z \\
 \end{array}} \right),
\]

\[
{\rm {\bf x}} \mapsto \left( {\begin{array}{l}
 x\cos \frac{\pi T_1 (z\sqrt 2 + 1)}{2} + y\sin \frac{\pi T_1 (z\sqrt 2 +
1)}{2},\,\,\, \\
 - x\sin \frac{\pi T_1 (z\sqrt 2 + 1)}{2} + y\cos \frac{\pi T_1 (z\sqrt 2 +
1)}{2},\,\, \\
 z \\
 \end{array}} \right),
\]

\[
{\rm {\bf x}} \mapsto \left( {\begin{array}{l}
 x,\,\,\, \\
 y\sqrt {\frac{z^2 + y^2}{z^2e^{ - 2\varepsilon (z^2 + y^2)} + y^2}} ,\,\,
\\
 z\sqrt {\frac{z^2 + y^2}{z^2 + y^2e^{2\varepsilon (z^2 + y^2)}}} \\
 \end{array}} \right),
\]

\begin{equation}
\label{eq6}
{\rm {\bf x}} \mapsto \left( {\begin{array}{l}
 x,\,\, \\
 y\cos \frac{\pi T_2 (x\sqrt 2 + 1)}{2} - z\sin \frac{\pi T_2 (x\sqrt 2 +
1)}{2},\, \\
 y\sin \frac{\pi T_2 (x\sqrt 2 + 1)}{2} + z\cos \frac{\pi T_2 (x\sqrt 2 +
1)}{2} \\
 \end{array}} \right).
\end{equation}

\section{Plykin attractor and method for validating the hyperbolicity}
\label{subsec:mylabel2}

We first set the fixed $T_{1}=T_{2}$=1, which, with a suitable choice of
$\varepsilon $, say, $\varepsilon = 0.77 $, ensures, according to [8],
occurrence of the Plykin attractor. Fig. 1 shows a phase portrait of the
attractor in axonometric projection obtained from results of
numerical calculations by iterations of the mapping
${\rm {\bf x}} \mapsto {\rm {\bf F(x)}}$.

\begin{figure}[htbp]
\begin{center}

\includegraphics[width=.5\textwidth,keepaspectratio]{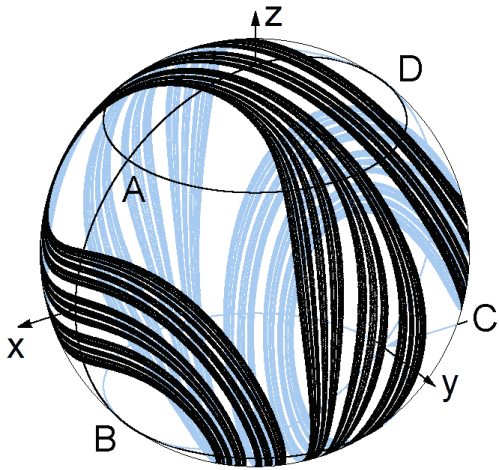}
\caption{Attractor of the mapping (\ref{eq6}) at $\varepsilon $=0.77 on the unit
sphere}
\label{fig1}
\end{center}
\end{figure}

The description of the dynamics can be reformulated in such way that the
instantaneous states of the system are represented by points on a plane. A
suitable replacement of the variables is
\begin{equation}
\label{eq7}
W = X + iY = \frac{x - z + iy\sqrt 2 }{x + z + \sqrt 2 },
\end{equation}
which corresponds to a stereographic projection from the sphere onto the
plane when taking a point C$( - 1 / \sqrt 2 ,\,\,0,\,\, - 1 / \sqrt 2 )$ as
the center of the projection. This point does not belong to the attractor
(it is in the ``hole''), therefore, the attractor obtained by the variable
change (\ref{eq7}) occupies a bounded area on the plane. The portrait of the
attractor on the plane is shown in Fig.2a. We note the presence of a
specific fractal transverse structure: the object looks like consisting of
stripes, each of which is composed of narrower stripes and so on.

\begin{figure}[htbp]
\includegraphics[width=.49\textwidth,keepaspectratio]{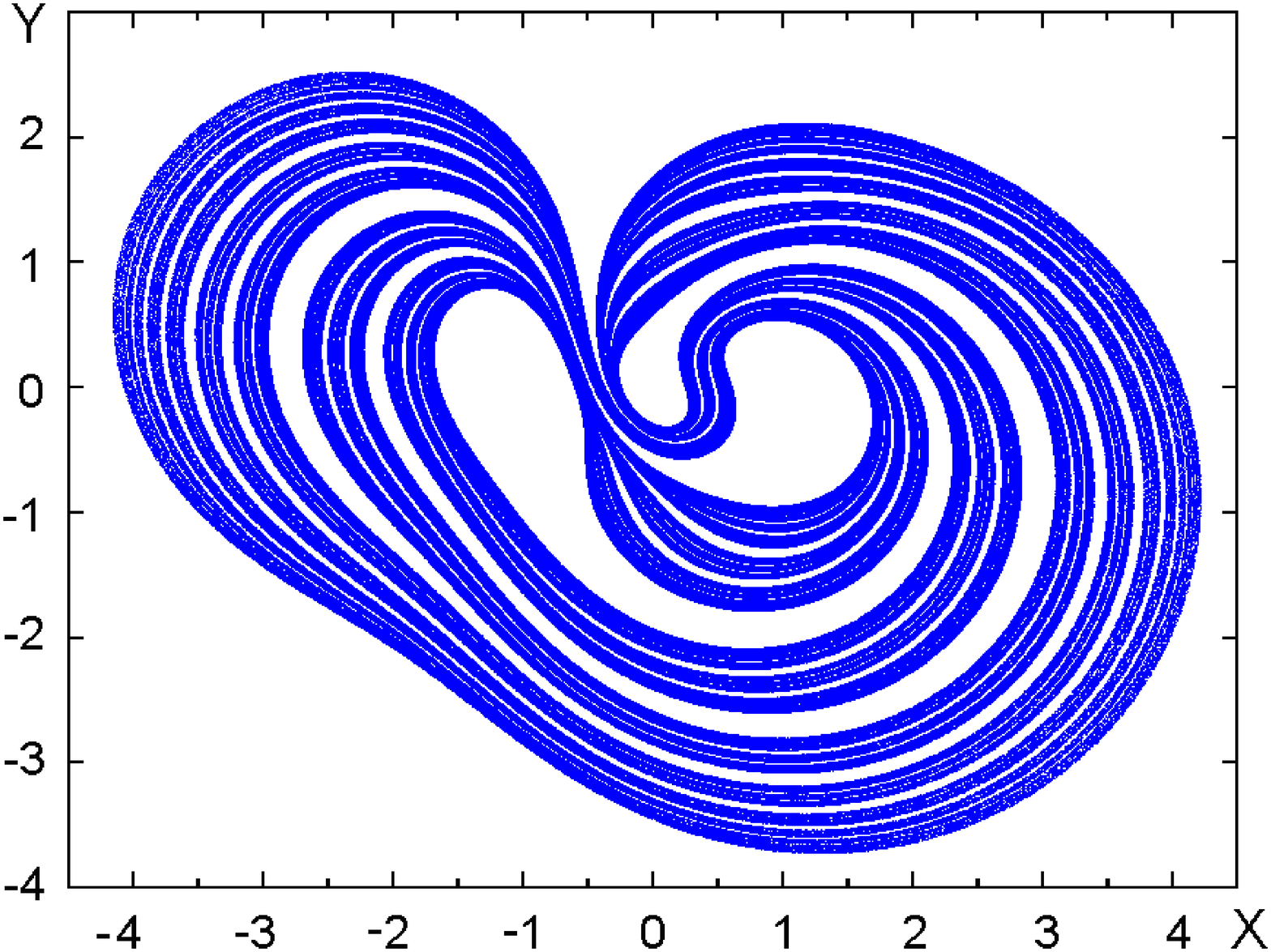}
\includegraphics[width=.49\textwidth,keepaspectratio]{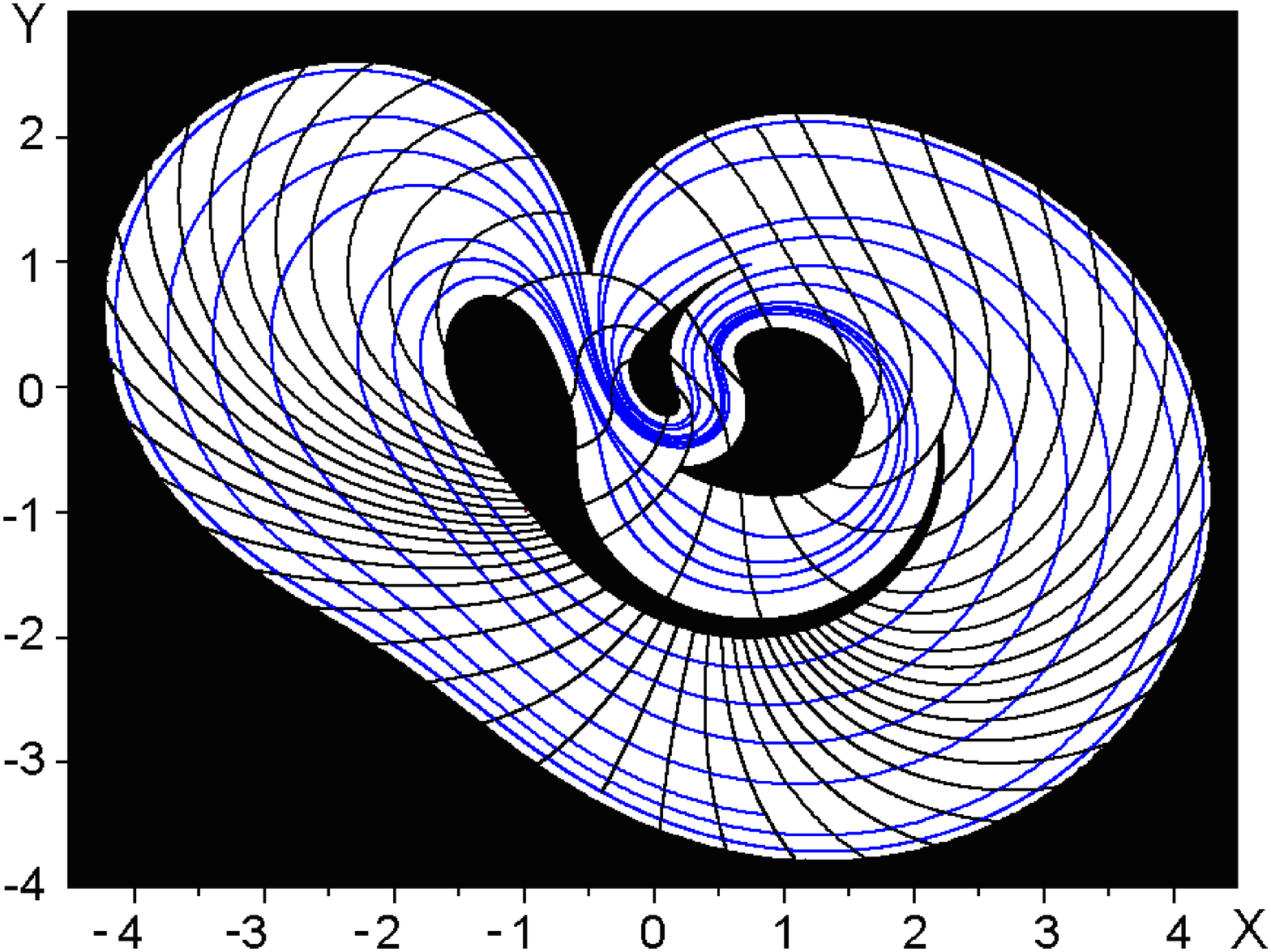}
\caption{Portrait of the attractor for the mapping ${\rm {\bf x}} \mapsto
{\rm {\bf F(x)}}$ at $\varepsilon $=0.77, $T_{1}$=$T_{2}$=1 depicted on the
plane (a). Stable and unstable foliations in the attractor absorbing area
shown in white, which are visualized from the results of computations (b).}
\label{fig2}
\end{figure}

To make sure that the attractor is hyperbolic, stable and unstable
foliations in the region containing the attractor are visualized in Fig.
2b.\footnote{ The construction method is as follows. First, for some point
{\bf x}, we find the image ${\rm {\bf \bar {x}}} = {\rm {\bf F}}^N({\rm {\bf x}})$
by iterations of the map and the inverse image by iterations of the inverse
map $\,{\rm {\bf \tilde {x}}} = {\rm {\bf F}}^{ - N}({\rm {\bf x}})$, where
N is an empirically selected integer. Then, using random initial conditions
in a small neighborhood of $\,{\rm {\bf \tilde {x}}}$, we carry out
iterations of the map and obtain a set of points $\,{\rm {\bf y}} = {\rm
{\bf F}}^N({\rm {\bf \tilde {y}}})$ that draw an unstable manifold. In the
same way, starting from the initial conditions ${\rm {\bf \bar {y}}}$ in a
small neighborhood of ${\rm {\bf \bar {x}}}$ and performing iterations of the
inverse mapping, we draw a stable manifold with points ${\rm {\bf y}} = {\rm
{\bf F}}^{ - N}({\rm {\bf \bar {y}}})$. The accuracy with which the graph is
obtained rapidly increases with $N$.} Stable and unstable manifolds are shown,
respectively, in black and blue. As seen from the figure, unstable manifolds
follow the attractor fibers, and stable manifolds are located across them.
The mutual disposition of the stable and unstable manifolds in the region
containing the attractor definitely excludes tangencies.

The Lyapunov exponents calculated for this attractor via the standard
algorithm, which uses iterations along the reference trajectory of two sets
of linearized equations accompanied with the Gram--Schmidt orthogonalization
[13], are

\begin{equation}
\label{eq8}
\Lambda _1 = 0.9587,\,\,\,\Lambda _2 = - 1.1407.
\end{equation}

The presence of a positive exponent indicates chaotic nature of the
attractor. The Kaplan--Yorke dimension of the attractor, which is determined
via the Lyapunov exponents, is $D_{KY} \approx $1.84. It serves as a
quantitative expression of the fractal structure of the bands forming the
attractor as observed in Figures 1 and~2.

One of the conventional computer tests for hyperbolicity consists in
calculating small perturbation vectors along a reference orbit on the
attractor during forward and backward evolution in time with evaluating the
angles between the vector subspaces responsible for instability [14,15,16].
If the statistical distribution of these angles is separated from zero
angles, then the dynamics is diagnosed as hyperbolic. In our case, the
subspaces of instability in direct time and in reverse time are
one-dimensional.

The procedure begins with calculating a rather long phase trajectory on the
attractor by iterations of the map. Further, the linearized equations are
solved forward in time along the reference trajectory with normalization of
the resulting vectors ${\rm {\bf a}}_n $ at the end of each switching
period. This vector determines the unstable direction at the points of the
orbit generated by the map. Further, along the same trajectory, linearized
equations are solved backward in time, from which the vectors ${\rm {\bf
b}}_n $ are obtained. Then, for each $n$, one calculates the angle between the
vectors ${\rm {\bf a}}_n $ and ${\rm {\bf b}}_n $, using the formula $\cos
\alpha _n = {\left| {{\rm {\bf b}}_n \cdot {\rm {\bf a}}_n } \right|}
\mathord{\left/ {\vphantom {{\left| {{\rm {\bf b}}_n \cdot {\rm {\bf a}}_n }
\right|} {\vert {\rm {\bf b}}_n \vert \vert {\rm {\bf a}}_n \vert }}}
\right. \kern-\nulldelimiterspace} {\vert {\rm {\bf b}}_n \vert \vert {\rm
{\bf a}}_n \vert }$.

Figure 3 shows a histogram of the angles obtained for system (\ref{eq6}) at
$\varepsilon $=0.77, $T_{1}=T_{2}=1$. It is clearly seen that the
distribution does not contain angles close to zero, which indicates the
hyperbolic nature of the attractor.

\begin{figure}[htbp]
\begin{center}
\includegraphics[width=.89\textwidth,keepaspectratio]{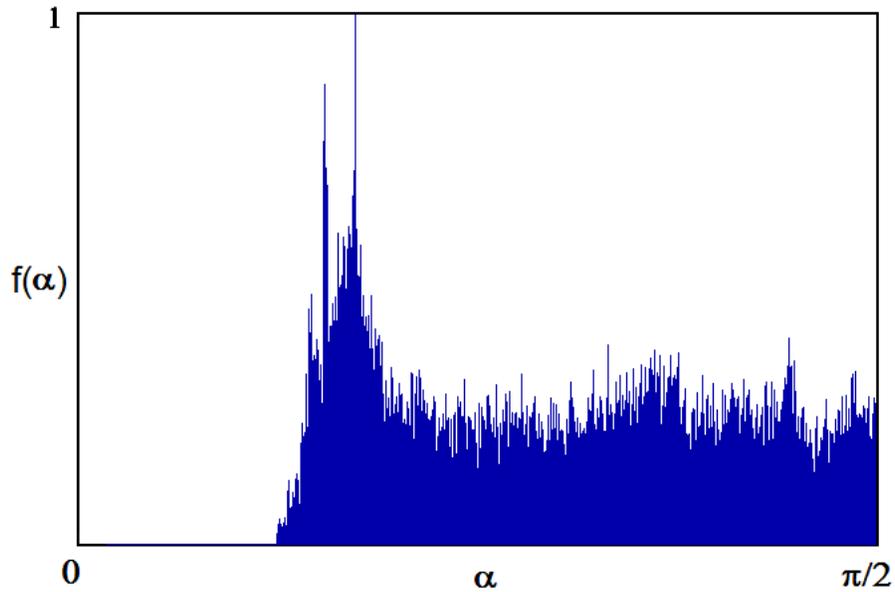}
\caption{Histogram of the distribution of angles $\alpha $ between the
subspaces of the perturbation vectors corresponding to instabilities in the
forward and reverse time along a reference orbit on the attractor,
obtained according to the method explained in the text.}
\label{fig3}
\end{center}
\end{figure}

\section{Charts of dynamical regimes. Chaotic and regular attractors}
\label{subsec:mylabel3}

\bigskip

To plot a chart of dynamical regimes, the parameter plane is scanned by
going through the grid nodes with a certain step in two parameters (see, for
example, [14, 18]). At each point, about $10^{3}$ iterations of the mapping
${\rm {\bf x}} \mapsto {\rm {\bf F(x)}}$ are performed, and according to the
results of the last iteration steps, an analysis is carried out for the
presence of a repetition period from 1 to 14 with some given level of
allowable error. If periodicity is detected, the corresponding pixel in the
diagram is indicated by a certain color, and the procedure continues with
analysis of the next point on the parameter plane. Similarly, the color
coding may be used not for the periods, but for the value of the largest
Lyapunov exponent; then we get a representation of the parameter plane as a
Lyapunov chart [19, 14]. Figure 4 demonstrates a chart of regimes on the
plane ($T_{1}$, $T_{2})$ for $\varepsilon $ = 0.77, on which, in addition to
periodic modes encoded with the corresponding colors, the areas of
quasiperiodicity (black color), chaos (white), and hyperbolic chaos (red
color, with the letter H) are shown.

\begin{figure}[htbp]
\begin{center}
\includegraphics[width=.99\textwidth,keepaspectratio]{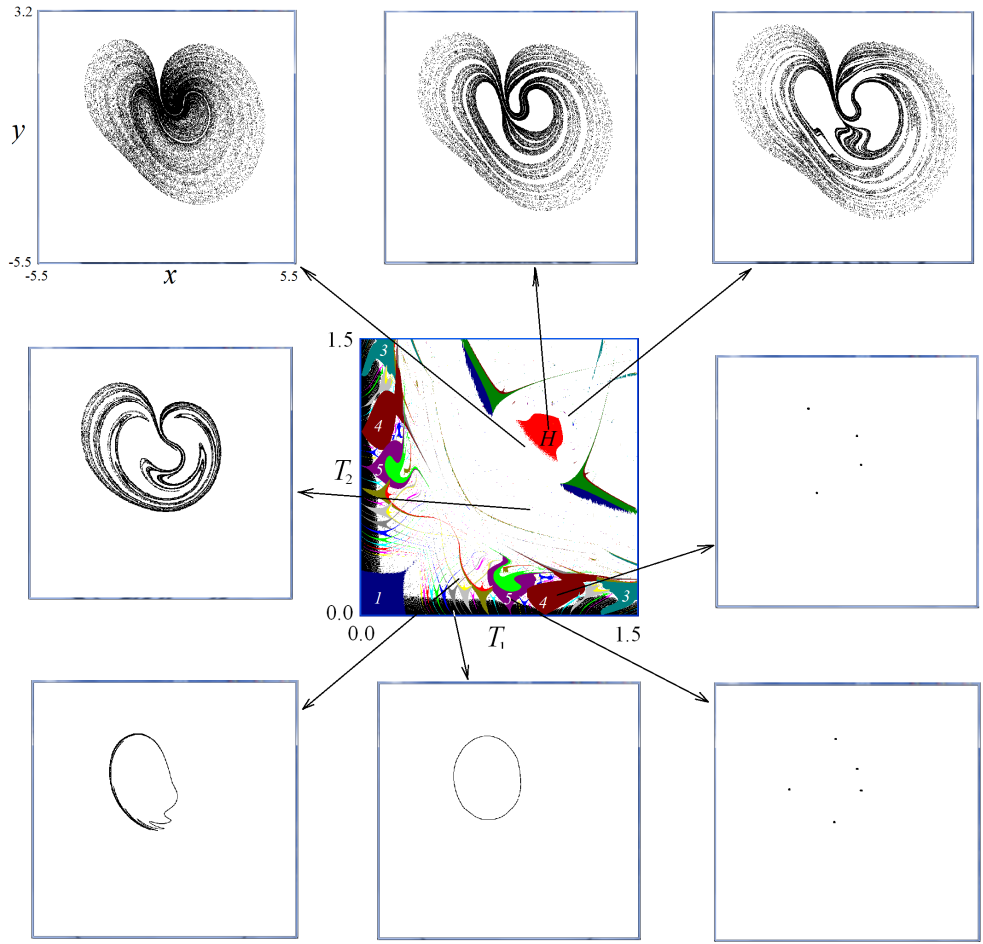}
\caption{In the center is the chart of regimes for $\varepsilon
=0.77$. Black corresponds to quasiperiodic modes, white corresponds to
chaotic ones, and other colors designate periodic modes. For the main
periodic tongues the periods are indicated by numbers. The region of
existence of the hyperbolic chaotic attractor is shown in red and marked
additionally with the letter H. The margin pictures show representative phase
portraits of attractors for some particular points on the parameter plane.}
\label{fig4}
\end{center}
\end{figure}

The chart is symmetrical with respect to the diagonal, i.e., with respect to
the replacement of the parameters $T_{1} \leftrightarrow T_{2}$, which is
natural given the structure of the equations (\ref{eq6}).

The hyperbolic attractor and periodic regimes occupy solid areas; this fact
is associated with their roughness.

Departure for a small distance in one or another direction from the region
of hyperbolic chaos leaves the attractor chaotic, but it becomes
non-hyperbolic.

Figure 5a shows graphs of the dependences of Lyapunov exponents on a
parameter counted along the main diagonal on the chart of Fig. 4. When
leaving the region of hyperbolicity in the diagonal direction up to the
right, the fibers forming the structure of the attractor undergo
deformation, but no visible broadening occurs, and the fractal band-in-band
structure is locally preserved. The Lyapunov exponents vary little in the
region of hyperbolicity, and the Kaplan--Yorke dimension remains greater
than 1 and noticeably less than 2. The graphs for the Lyapunov dimension
along the diagonal are shown in Figure 5b.

\begin{figure}[htbp]
\begin{center}
\includegraphics[width=.89\textwidth,keepaspectratio]{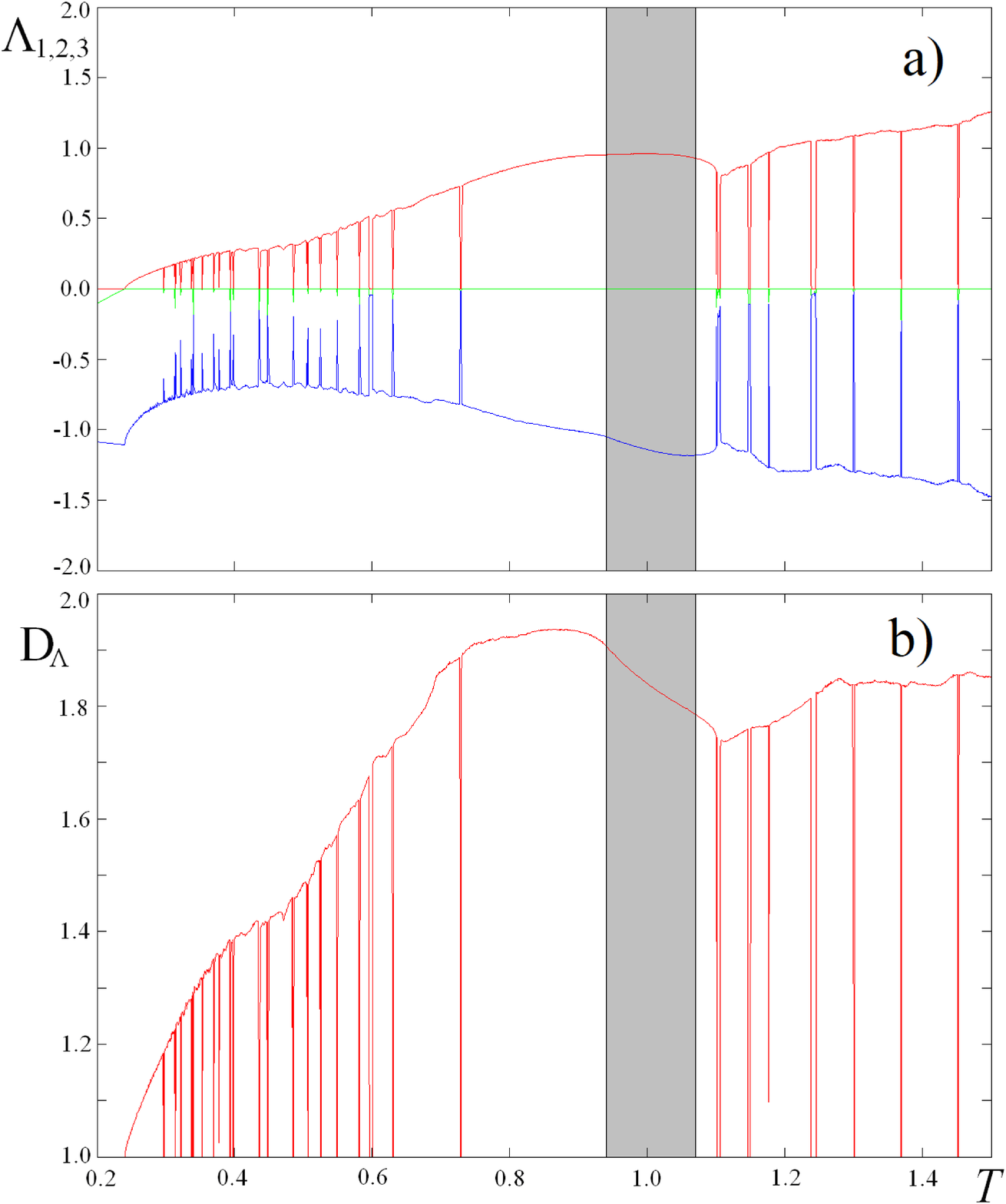}
\caption{Dependencies of the Lyapunov exponents (a) and of the Lyapunov
dimension (b) on the parameter $T=T_{1}=T_{2}$, which is counted along the
main diagonal on the chart of Fig. 4. Gray tone indicates the region of
hyperbolicity of the attractor.}
\label{fig5}
\end{center}
\end{figure}

When going diagonally to the left down, the situation looks different,
namely, as ``closing the holes'', which were ensuring the existence of the
Plykin attractor, while the bands that make up the structure of the
attractor undergo broadening, close together, merge, and the fractal
structure fade away. The fractality disappearance is also evidenced by the
fact that the two Lyapunov exponents become close in absolute value, which
means that the Kaplan--Yorke dimension of the attractor is approaching 2.

On the parameter plane, in a strip along the vertical axis (small $T_{1}$
values) and in a strip along the horizontal axis (small $T_{2}$ values), i.e.
where the duration of one stage of the differential rotation is much shorter
than the other, one can see a characteristic picture of the Arnold
synchronization tongues [20, 14] corresponding to periodic regimes of
different periods. Between tongues there are areas of quasiperiodic
dynamics. An exit from each tongue along the path inside it in the direction of
expanding the tongue (i.e., up for the set of tongues in the bottom part of
the chart, or right for the tongues in the left part of the diagram) is
accompanied by period doubling bifurcations with transition to
non-hyperbolic chaos according to the Feigenbaum scenario [21,14].

The regions of non-hyperbolic chaos between the Arnold tongues and the
Plykin attractor have inclusions in a form of ``shrimps'' [22,23], where
periodic dynamics is restored. This is characteristic of a fragile chaos
situation associated with the mathematical concept of a quasiattractor [3],
which occur in many other chaotic systems studied in the literature (H\'enon
map, R\"{o}ssler model, etc.).

Figures 6 and 7 show charts for other two-dimensional sections of the
three-dimensional parameter space of the system (\ref{eq6}). Inspection of these
charts shows that the region of the hyperbolic attractor is bounded from
above in the parameter $\varepsilon $. With increase of
$\varepsilon $, the system manifests a developed multistability.

\begin{figure}[htbp]
\begin{center}
\includegraphics[width=.99\textwidth,keepaspectratio]{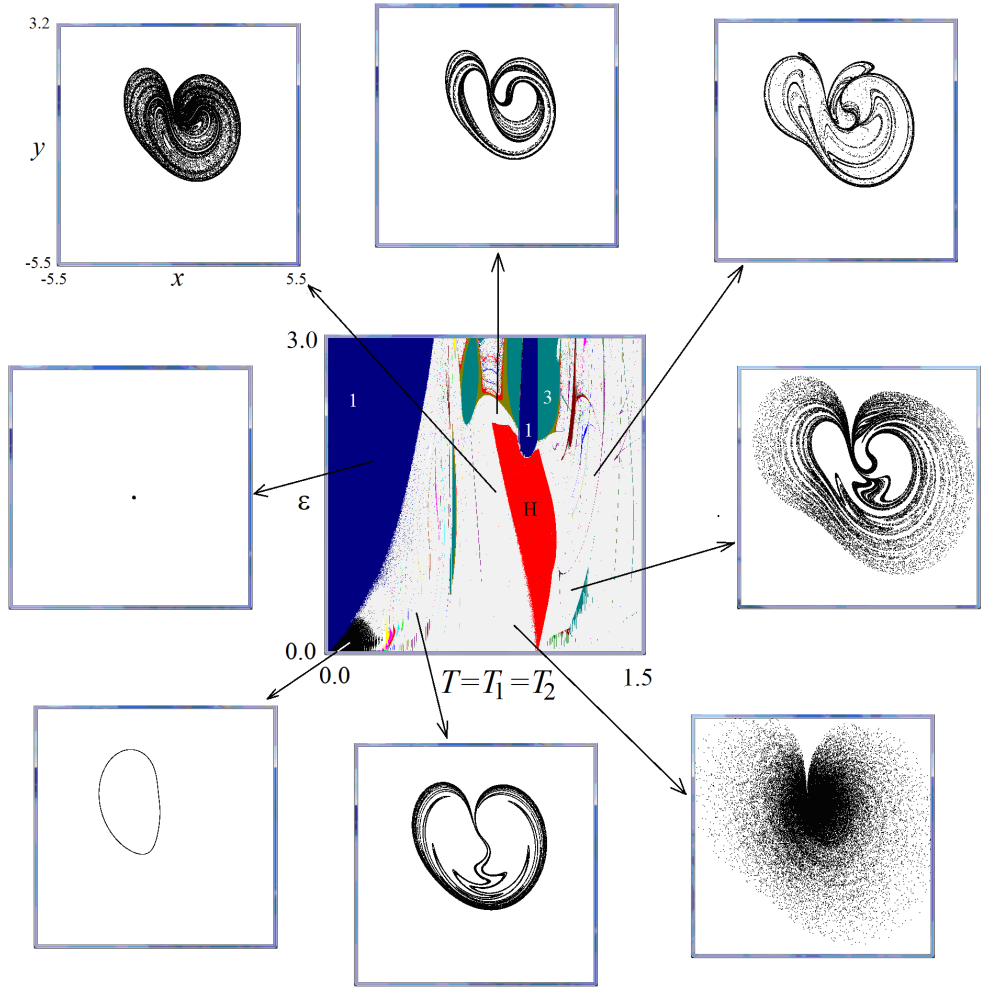}
\caption{In the center is the chart of regimes on the parameter plane
($T=T_{1}=T_{2}$, $\varepsilon )$. Black corresponds to quasiperiodic
modes, white corresponds to chaotic ones, and other colors designate
periodic modes. For the main periodic tongues the periods are indicated by
numbers. The region of existence of the hyperbolic chaotic attractor is
shown in red and marked with the letter H. The margin pictures  show
representative phase portraits of attractors for some particular points on
the parameter plane.}
\label{fig6}
\end{center}
\end{figure}

\begin{figure}[htbp]
\begin{center}
\includegraphics[width=.99\textwidth,keepaspectratio]{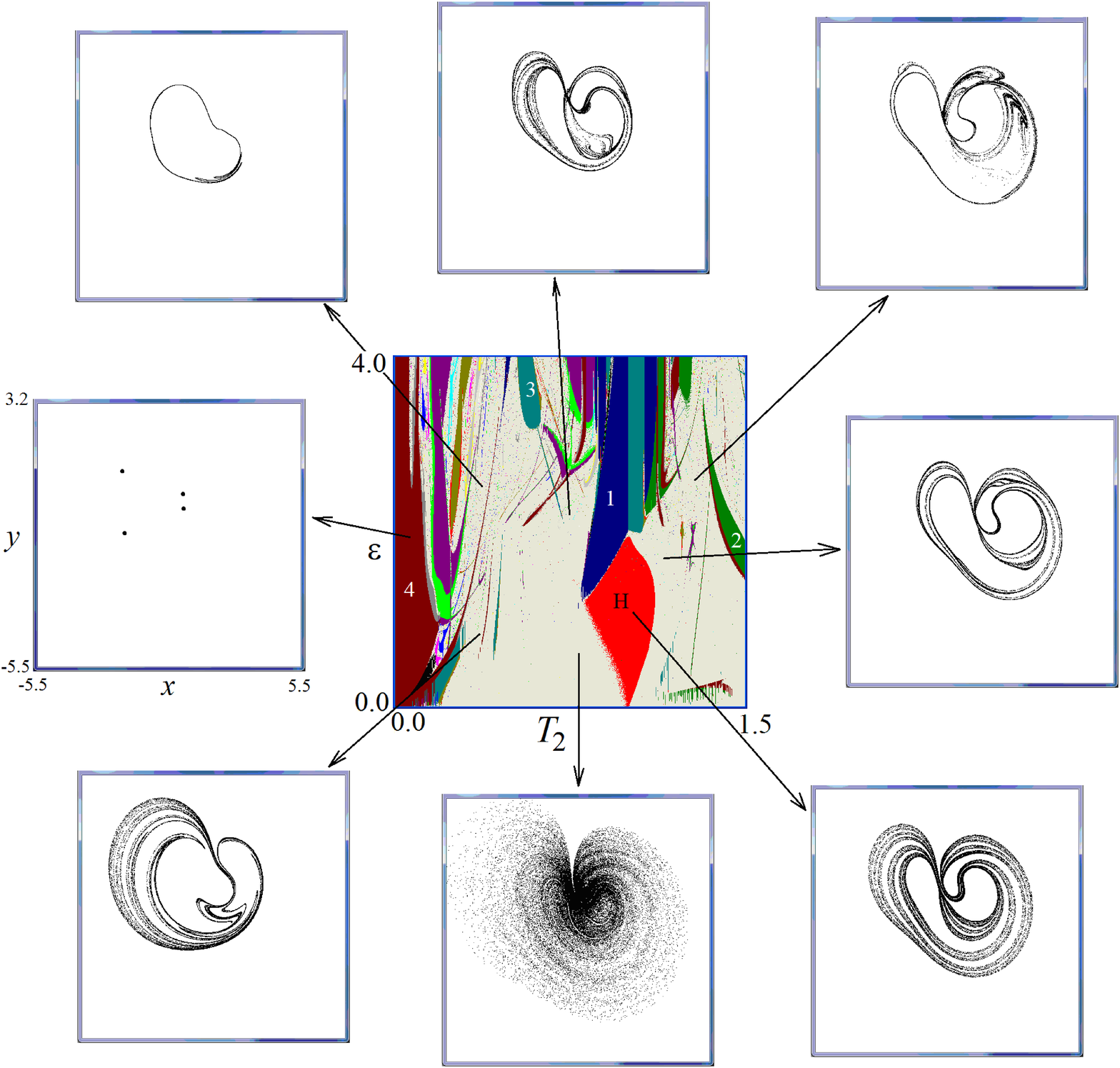}
\caption{In the center is the chart of regimes on the parameter plane
($T_{2}$, $\varepsilon )$ at $T_{1}=1$. Black corresponds to quasiperiodic
modes, white corresponds to chaotic ones, and other colors designate
periodic modes. For the main periodic tongues the periods are indicated by
numbers. The region of existence of the hyperbolic chaotic attractor is
shown in red and marked with the letter H. The margin pictures show
representative phase portraits of attractors for some particular points on
the parameter plane.}
\label{fig7}
\end{center}
\end{figure}

\section{Conclusions}
\label{subsec:mylabel4}

We have considered a model system with Plykin type hyperbolic chaotic
attractor in a map resulting from several successive transformations of a
sphere such as differential rotations around two orthogonal axes alternating
with stages of dissipative evolution. The attractor can also be considered
on a plane by applying a stereographic projection.

The dependence of the dynamical behavior of the system on parameters
specifying the duration of the stages of differential rotation and the
dissipation parameter is discussed. Charts of dynamical regimes are
presented, which make it possible to judge about mutual location of the
regions of periodic, quasiperiodic, chaotic, and hyperbolic chaotic behavior
in the parameter space of the system.

It is shown that the hyperbolic attractor and the periodic modes occupy
solid regions in the parameter space, which is obviously due to the
roughness of the corresponding types of dynamical behavior.

Upon leaving the periodicity regions, scenarios of transition to
quasiperiodicity and chaos known in nonlinear dynamics occur, which
correspond to the structures in the form of Arnold tongues in the parameter
space. When leaving the hyperbolic chaos region of for a short distance, the
attractor remains chaotic, but becomes non-hyperbolic. Two scenarios stand
out qualitatively depending on the direction of the exit. One corresponds to
a situation that the fibers forming the attractor structure undergo
deformation, but without their visible broadening, and the ``strip in
strip'' fractal structure is locally preserved. The second scenario looks
like ``closing the holes'' which were ensuring existence of the Plykin
attractor, when the bands that make up the structure of the attractor
undergo broadening, and the attractor loses its
fractal structure. The region of non-hyperbolic chaos manifests inclusions
in the form of ``shrimps'', where periodic dynamics is restored that is
typical for non-rough chaos.

Summarizing, the first qualitative information on the parameter space
arrangement around the region of the Plykin attractor has been obtained. It
is of obvious interest to continue researches in this direction to identify
details of the bifurcation transitions, to construct a complete
classification of the scenarios of birth and destruction of hyperbolic
chaos, and to develop comparative analysis of the pictures of the parameter
spaces near the regions of occurrence of different types of uniformly
hyperbolic attractors.

\section*{Acknowledgement}
\label{subsec:acknowledgemente}

\textit{The work is supported by Russian Science Foundation, Grant No 17-12-01008.}


\end{document}